\documentclass{andp2012}
\usepackage[english]{babel}
\usepackage{bm}
\usepackage{mathtools}
\newcommand{\be}{\begin{equation}}
\newcommand{\ee}{\end{equation}}

\newcommand{\im}[1]{\text{Im}[{#1}]}


\title{Anomalous minimum and scaling behavior of localization length near an isolated flat band}

\author[F. Author]{L. Ge\inst{1,2}\footnote{E-mail:~\textsf{li.ge@csi.cuny.edu}}}
\address[1]{Department of Engineering Science and Physics, College of Staten Island, CUNY, Staten Island, NY 10314, USA}
\address[2]{The Graduate Center, CUNY, New York, NY 10016, USA}

\begin{abstract}
Using one-dimensional tight-binding lattices and an analytical expression based on the Green's matrix, we show that anomalous minimum of the localization length near an isolated flat band, previously found for evanescent waves in a defect-free photonic crystal waveguide, is a generic feature and exists in the Anderson regime as well, i.e., in the presence of disorder. Our finding reveals a scaling behavior of the localization length in terms of the disorder strength, as well as a summation rule of the inverse localization length in terms of the density of states in different bands. Most interestingly, the latter indicates the possibility of having two localization minima inside a band gap, if this band gap is formed by two flat bands such as in a double-sided Lieb lattice.
\end{abstract}
\shortabstract

\begin{document}
\maketitle

\pagestyle{empty}
\section{Introduction}

Systems that exhibit flat bands have attracted considerable interest in the past decades, including optical \cite{Manninen1,Manninen2} and photonic lattices \cite{Vicencio,Mukherjee,Rechtsman,Silva,flatbandPT}, graphene \cite{Kane,Guinea},  superconductors \cite{Simon,Kohler1,Kohler2,Imada}, fractional quantum Hall systems \cite{Tang,Neupert,Sarma} and exciton-polariton condensates \cite{Jacqmin,Baboux}. As the name suggests, a flat band is dispersionless, or in other words, its density of states (DOS) diverges at a particular energy, known as the flat band energy. A quantity strongly influenced by DOS is the the localization length in a disordered system, and it displays quite different properties depending on whether a flat band is present. For example, previous studies in flat band systems have shown inverse Anderson transition \cite{Goda}, localization with unconventional critical exponents and multi-fractal behavior \cite{Chalker}, mobility edges with algebraic singularities~\cite{Bodyfelt}, and unusual scaling behaviors \cite{Flach,Leykam}.

Despite all these interesting findings, there is one important aspect that has not been investigated systematically, i.e.,  the energy dependence of the localization length near an isolated flat band, which is separated from its neighboring band on each site by a band gap. To be clear, here by localization we refer to the finite spatial extension of the wave function. There are two mechanisms typical in a periodic system that can lead to wave function confinement, one by evanescent waves inside a band gap \cite{PhC_book}, where the DOS (of the propagating states) vanishes. The other is Anderson localization \cite{Anderson}, where disorder, ubiquitous in naturally formed and fabricated materials, suppresses wave diffusion.

One usually finds a minimum of the localization length near the center of a band gap (``midgap"), where it is most unlikely to couple to the neighboring (propagating) bands \cite{Urbach,Cody,Mott_book,Cohen,Economou}. An exception to this conventional understanding was reported in a photonic crystal waveguide \cite{PhC1}, formed by removing one row of unit cells from a two-dimensional (2D) periodic dielectric slab. It was found that the shortest evanescent tail in the transverse direction, perpendicular to the removed row, is at a frequency shifted significantly away from the midgap. This observation was later explained using the complex wave vector inside the band gap of the perfect 2D slab (before the row removal) \cite{PhC2}, where a flat band was found on one side of the band gap as the result of a Van Hove saddle point singularity.

In this work we show that these anomalous localization minima are a generic feature of an isolated flat band and independence of the origin of the latter. Hence these anomalous minima exist in the Anderson regime as well, i.e., in the presence of disorder. We exemplify this finding using a one-dimensional (1D) tight-binding Lieb lattice, and we attribute these anomalous minima to the compact states in the flat band \cite{Manninen2,Baboux}, which is supported by an analytical expression of the inverse localization length. This expression reveals a scaling behavior of the localization length in terms of the disorder strength, as well as a summation rule of the inverse localization length using an integral transform of the DOS in the flat band and the opposing band. We find the latter term to be almost a constant near the flat band energy (denoted by $E_{FB}$ below) for weak disorder, and the flat band term shows a non-monotonic energy dependence that leads to the anomalous minima of the localization length. Most interestingly, we show that the localization length can exhibit \textit{two} minima instead of one inside a band gap, if this band gap is formed by two flat bands such as in a double-sided Lieb lattice.

\begin{figure}[t]
\includegraphics[clip,width=\linewidth]{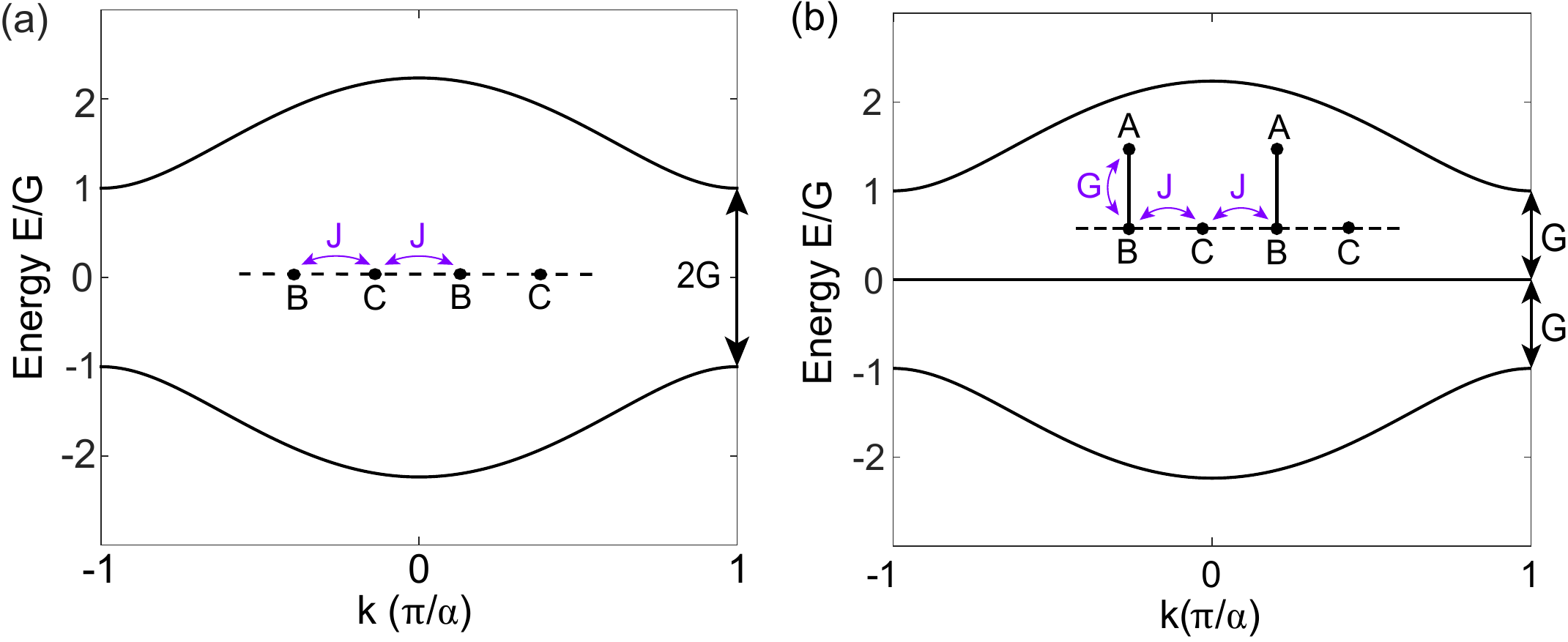}
\caption{Band structure of a simple periodic lattice (a) and a Lieb lattice (b). In (a) $E_B=G$, $E_C=-G$, and $G=1$. In (b) $E_{A}=E_{B}=E_{C}=0$, $G=J=1$, and the flat band is located at $E_{FB}=0$. $\alpha$ is the lattice constant.}\label{fig:bands}
\end{figure}

\section{Properties of the localization length}
\subsection{Evanescent waves in the absence of disorder}
Before discussing the anomalous localization minima in the presence of disorder, we first briefly review the properties of evanescent waves inside a band gap in the absence of disorder. A band gap, by definition, does not support any extended or propagating states. In other words, the wave vector inside a band gap takes a complex value, and its imaginary part gives the inverse localization length of the evanescent wave \cite{PhC_book}. From this point of view, two systems with identical band structures (except at some singular values of the energy), should also have identical localization properties for evanescent waves inside the band gap.

As a case in point, we compare a simple 1D periodic lattice and a Lieb lattice in Fig.~\ref{fig:bands}. They have two identical dispersive bands given by
\be
E = \pm\sqrt{G^2 + 2J^2(1+\cos{k\alpha})},\label{eq:dispbands}
\ee
with the Lieb lattice featuring an additional flat band at $E_{FB}=0$. These band structures are calculated using a tight-binding model, with nearest neighbor couplings $J$ and $G$ shown in the insets of Fig.~\ref{fig:bands}. Take the Lieb lattice for example, the tight-binding model is given by
\begin{align}
i\dot{A}_j &= E_{Aj}A_j + G B_j,\label{eq:A0}\\
i\dot{B}_j &= E_{Bj}B_j + G A_j + J(C_j+C_{j-1}),\label{eq:B0}\\
i\dot{C}_j &= E_{Cj}C_j + J (B_j + B_{j+1}),\label{eq:C0}
\end{align}
where the overhead dot denotes the time derivative and we have taken $\hbar=1$. $X_j$ is the value of the wave function on the $j$th site of type $X\,(X=A,B,C; j=1,2,\ldots)$, and $E_{Xj}$ is the energy on site $X_j$. Clearly the wave function and hence the localization length do not depend on the overall energy scale, because all the terms in Eqs.~(\ref{eq:A0})--(\ref{eq:C0}) are linear in energy. We can thus use an energy scale at will, which we choose to be $G$ and set it to be 1.

The model above assumes noninteracting particles, with which the localization of fermions, bosons, and classical waves can be treated on the same footing in their respective regime of validity, i.e., as a general wave phenomenon where propagation is suppressed \cite{review}. We also note that although Eqs. (\ref{eq:A0})--(\ref{eq:C0}) are written in the form of the Schr\"odinger equation, it applies to the wave equation as well in the paraxial regime, which can be realized, for example, by coupled parallel waveguides \cite{Rechtsman,Vicencio}. There the time derive is replaced by the spatial derivative along the axial (propagation) direction, and the energy is replaced by the propagation constant \cite{segev}.

To find the localization length of the evanescent wave inside the band gap(s) shown in Fig.~\ref{fig:bands}, we simply need to invert Eq.~(\ref{eq:dispbands}) and solve $k$ as a function of $E$:
\be
k(E) = \pi \pm i\ln\left(\sigma(E)+\sqrt{\sigma^2(E)-1}\right),
\ee
where $\sigma(E)\equiv 1 + (G^2-E^2)/2J^2$.
This result holds for both the simple periodic lattice and the Lieb lattice in $E\in[-G,G]$, except right at the flat band energy $E_{FB}=0$ for the Lieb lattice where Eq.~(\ref{eq:dispbands}) does not apply. Away from this singular point, the localization length of the evanescent wave for both lattices is then given by
\be
\xi = |\im{k}|^{-1} = \left[\ln\left(\sigma+\sqrt{\sigma^2-1}\right)\right]^{-1}. \label{eq:xi_complexk}
\ee
This localization length can be probed by sending a plane wave at energy $E$ into the system or by introducing a \textit{point} defect on the lattice. The latter approach was used in Ref.~\citenum{flatbandPT} where $E_A$ in one unit cell of the Lieb lattice is detuned by $G/10$, and the resulting defect state confirms an evanescent tail given by Eq.~(\ref{eq:xi_complexk}). One can verify in the same way that Eq.~(\ref{eq:xi_complexk}) holds for the evanescent wave in the simple periodic lattice as well, similar in spirit to introducing a row defect in a 2D square lattice \cite{PhC1,PhC2}.

Importantly, we note that the localization length given by Eq.~(\ref{eq:xi_complexk}) has a minimum at the midgap ($E=E_{FB}$), which follows the conventional understanding mentioned in the introduction. Although using the approach above we cannot evaluate the localization length at $E_{FB}$ on the Lieb lattice, it is clear that the presence of the flat band will no doubt make the localization length discontinuous at $E_{FB}$. This observation can be thought as the precursor of the anomalous localization minimum we will discuss in the next section, where disorder broadens the flat band and smooths out the discontinuity of the localization length.

\subsection{With disorder}

\begin{figure}[b]
\includegraphics[clip,width=\linewidth]{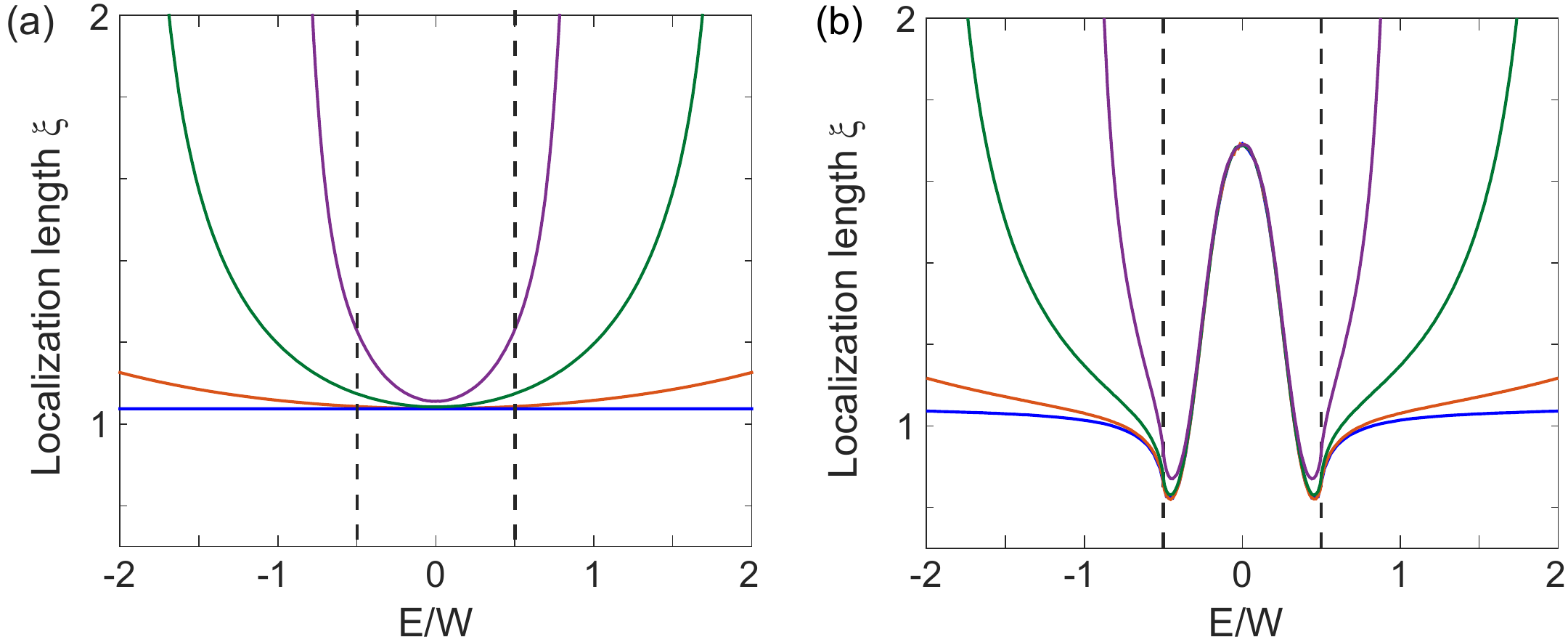}
\caption{Localization length of the lattices in Fig.~\ref{fig:bands} in the presence of disorder. The vertical dashed lines show the boundaries of the on-site energy disorder ($E\in[-W/2,W/2]$), and the localization length is calculated with $W=0.01,0.2,0.5$ and 1 (from bottom to top).}\label{fig:loclen}
\end{figure}

The localization length $\xi$ with disorder can be calculated straightforwardly using the transfer matrix method (see Appendix \ref{sec:Tmatrix}).
Below we assume a uniform disorder $W$, with $E_{Xj}\in [E_X-W/2,E_X+W/2]$ and $\langle E_{Xj}^2\rangle=W^2/12$.
The results are shown in Fig.~\ref{fig:loclen} as a function of the scaled energy $u\equiv(E-E_{FB})/W$ and in units of the lattice constant, with different values of $W$.

While $\xi$ of the simple periodic lattice behaves similarly to Eq.~(\ref{eq:xi_complexk}) and displays a minimum at $E=0$, we  note that for a given weak disorder $W\lesssim G$, the localization length $\xi(u;W)$ of the Lieb lattice exhibits a minimum in each of the two band gaps formed by the flat band and its neighboring dispersive bands (i.e., $u\in[-G/W,0]$ and $[0,G/W]$). In addition, $\xi(u;W)$ barely changes with the disorder strength $W$ for $|u|\lesssim1/2$, hence displaying a scaling behavior in terms of $u$.


Importantly, the localization minima at $|E|\lesssim W/2$ [see Fig.~\ref{fig:loclen}(b)] are far from the respective midgap at $E=\pm G/2$ when $W\ll G$, to which we refer as the anomalous localization minima.
To further distinguish them from the conventional midgap ones as shown in Fig.~\ref{fig:loclen}(a), we note that here the curvature of the localization length changes sign near $|E|=W/2$.

\begin{figure}[h]
\includegraphics[clip,width=\linewidth]{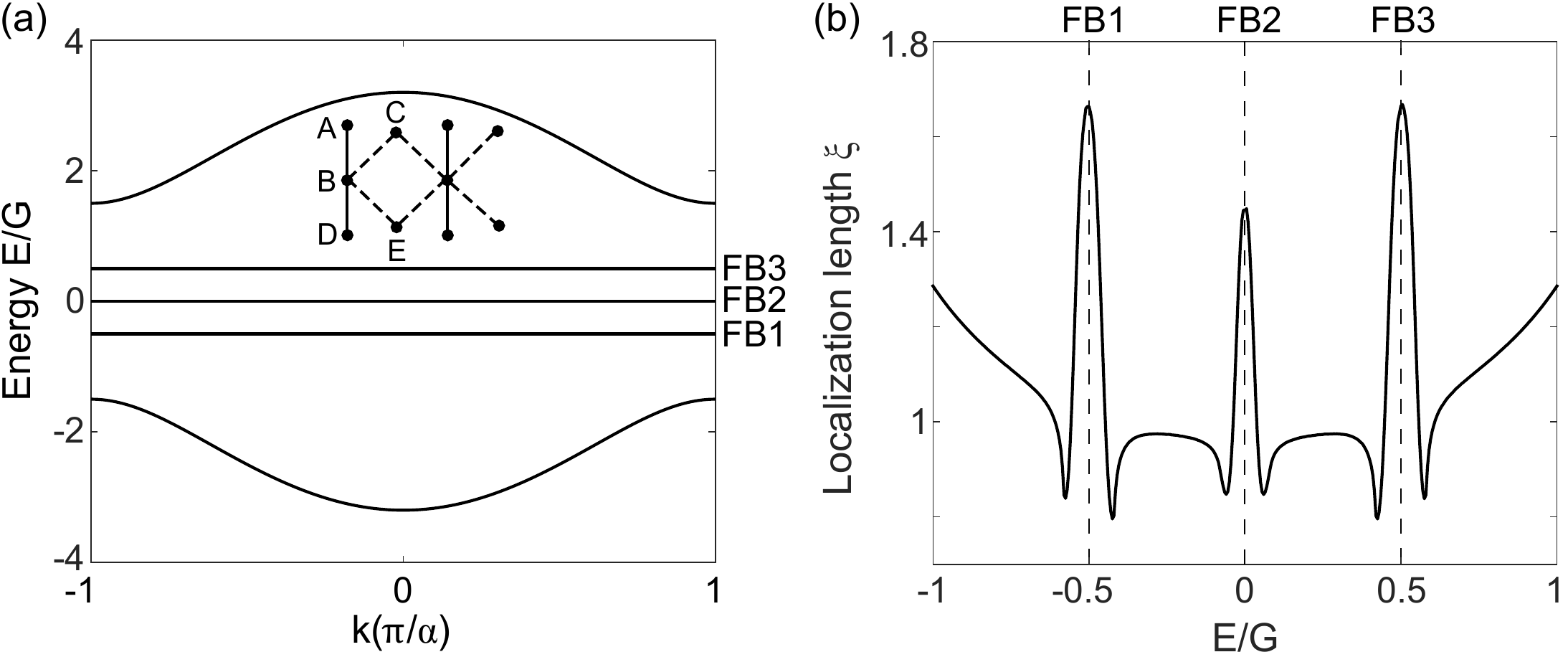}
\caption{ Band structure (a) and localization length (b) of a 1D double-sided Lieb lattice. Inset in (a): Schematic of the lattice structure. Solid and dashed lines indicate couplings $G$ and $J$, respectively. Three flat band exist at $E_{FB1}=-0.5$, $E_{FB2}=0$ and $E_{FB3}=0.5$, and we have taken $E_{A}=E_{C}=0.5$, $E_{D}=E_{E}=-0.5$, $E_B=0$ and $G=J=1$.  The localization length shown in (b) is calculated with $W=1/6$ using the transfer matrix method, and the dashed lines mark the flat band energies.}\label{fig:2minima}
\end{figure}

As we will elucidate in the next section, these anomalous minima are caused solely by the flat band and independent of the dispersive bands on the other side of the band gaps. Here to highlight the anomaly of these localization minima, we provide another example using a double-sided Lieb lattice shown in Fig.~\ref{fig:2minima}. This lattice has three neighboring flat bands, and the localization length exhibits \textit{two minima} (instead of one) inside \textit{each} gap formed by two flat bands, which shows clearly that each minimum is due to the closest flat band.

\vspace{-1.5cm}
\section{Localization length and DOS}

\begin{figure}[t]
\includegraphics[clip,width=\linewidth]{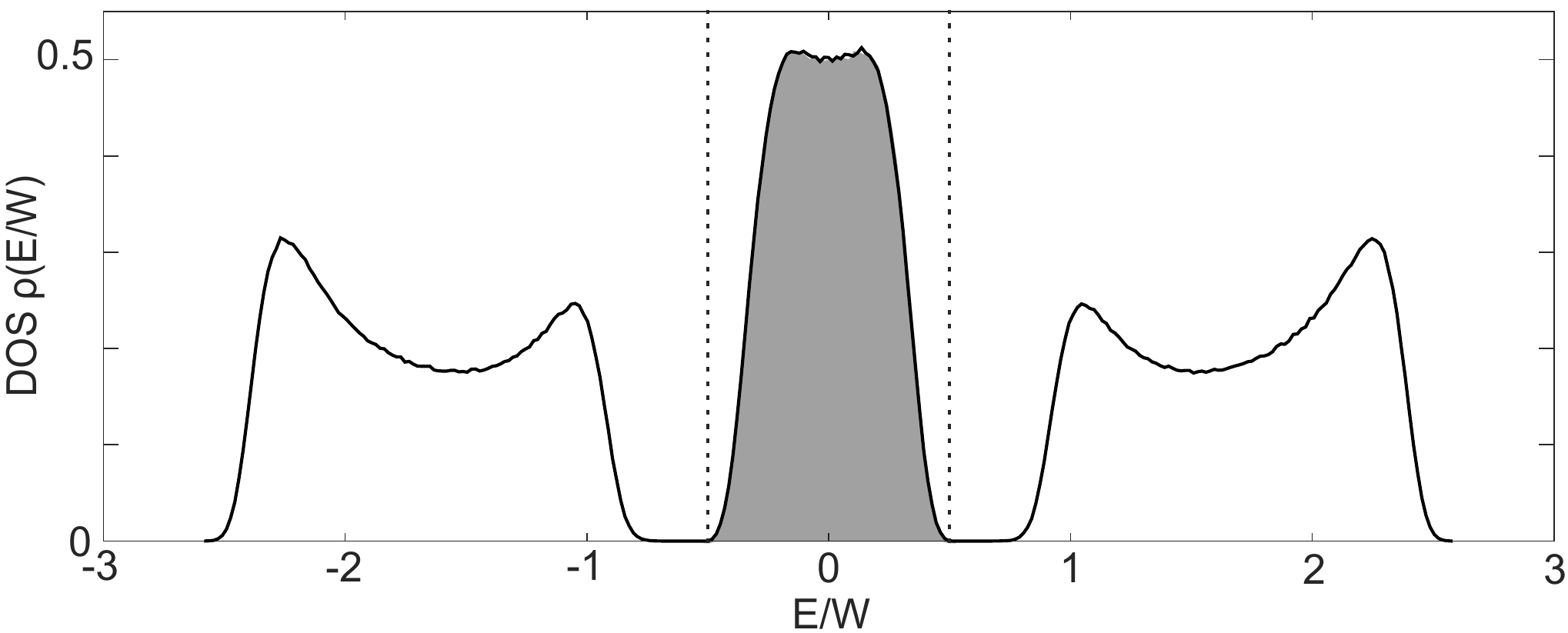}
\caption{DOS of the Lieb lattice shown in Fig.~\ref{fig:bands}(b) with $W=1$ (solid line). Its value for the flat band with $W=0.5$ is shown by the shadowed area, and the dotted lines mark $E=\pm W/2$. 100 unit cells and $10,000$ samples are used.}\label{fig:dos}
\end{figure}

To prove that the anomalous minima of the localization length in Fig.~\ref{fig:loclen}(b) are solely caused by the flat band, we use an integral transform of the DOS that gives the inverse localization length \cite{Herbert,Thouless}. This approach utilizes the top right element of the Green's matrix $\bm{\mathcal{G}}$ that represents the probability amplitude of wave propagation from one end of a finite lattice with $N$ unit cells to the other end. For the 1D Lieb lattice, we find this matrix element to be
\be
\mathcal{G}_{1,3N} 
= \frac{\text{cofactor}(E-\bm{H})_{3N,1}}{\det(E-\bm{H})}
=\frac{GJ^{2N-1}\prod_{j=2}^N \Delta_{Aj}}{\prod_{k=1}^{3N}(E-\lambda_{k})}, \label{eq:GreenMatrix}
\ee
where $\bm{H}$ is the $3N\times3N$ tight-binding Hamiltonian represented by Eqs.~(\ref{eq:A0})--(\ref{eq:C0}), $\{\lambda_{k}\}$ are its $3N$ eigenvalues, and ``$\text{cofactor}$" denotes the cofactor matrix. The localization length $\xi(E)$ in this approach is given by
\be
\xi^{-1}(E) = -\lim_{N\rightarrow\infty} \frac{1}{N} \ln |\mathcal{R}_{1,3N}|,
\ee
where the dimensionless quantity $\mathcal{R}_{1,3N}$ is the residual of $\mathcal{G}_{1,3N}$ at the eigenvalue $\lambda_k$ closest to $E$. Using Eq.~(\ref{eq:GreenMatrix}), we derive a summation rule for the inverse localization length, i.e.,
\be
\xi^{-1}(E;W) = \xi^{-1}_{FB}(u) + \xi^{-1}_{DB}(E;W), \label{eq:summation}
\ee
where
\begin{gather}
\xi_{FB}^{-1}(u) = \hspace{-1mm}\int_{-\frac{1}{2}}^{\frac{1}{2}}\,\ln\left|u-v\right|[\rho_{FB}(v)-1]\,dv \label{eq:dos_lieb3},\\
\xi_{DB}^{-1}(E;W) = 2\int_{-\infty}^\infty\hspace{-2mm}\ln\left|\frac{E-z}{J}\right|\,\rho_{DB}(z;W)\,dz \label{eq:dos}
\end{gather}
are the contributions of the flat band and dispersive bands, respectively. $\rho_{FB}(v),\rho_{DB}(z;W)$ are their respective DOS normalized by $\int_{-\infty}^{\infty} \rho_{FB}(v)\,dv=1$ and $\int_{-\infty}^{\infty} \rho_{DB}(z;W)$ $dz=1$. They are related to the total DOS $\rho(z;W)$ (see Fig.~\ref{fig:dos}) by $\rho_{FB}(v)=3W\rho(z;W)\,(v\equiv z/W\in[-1/2,1/2])$ and $\rho_{DB}(z;W)=1.5\rho(z;W)\,(|z|>W/2)$, so defined because there are twice as many states in the dispersive bands as in the flat band and we have used $\int_{-\infty}^{\infty} \rho(z;W)\,dz=1$. Equation~(\ref{eq:summation}) gives a good agreement with the results of the transfer matrix approach (see Appendix \ref{sec:compare}).

Note that we have taken (1) $E=E_{FB}\pm W/2$ to be the boundaries between the broadened flat band and dispersive bands; and (2) $\rho_{FB}(v)$ to be independent of the disorder strength $W$, which are good approximations for weak disorder as exemplified by Fig.~\ref{fig:dos}. The former can be understood since $\pm W/2$ is the maximum perturbation on each lattice site, and the latter can be understood from the perspective of degenerate perturbation theory: the flat band states are degenerate on the clean lattice, and to the leading order the perturbed flat band states are simply their linear superpositions that are orthogonal with respect to the disorder potential. Clearly multiplying a constant to the disorder potential does not change the orthogonality of these superpositions, and hence the energy of the broadened flat band states scales linearly with the perturbation $W$, as long as the higher-order perturbations can be neglected.

\begin{figure}[b]
\includegraphics[clip,width=\linewidth]{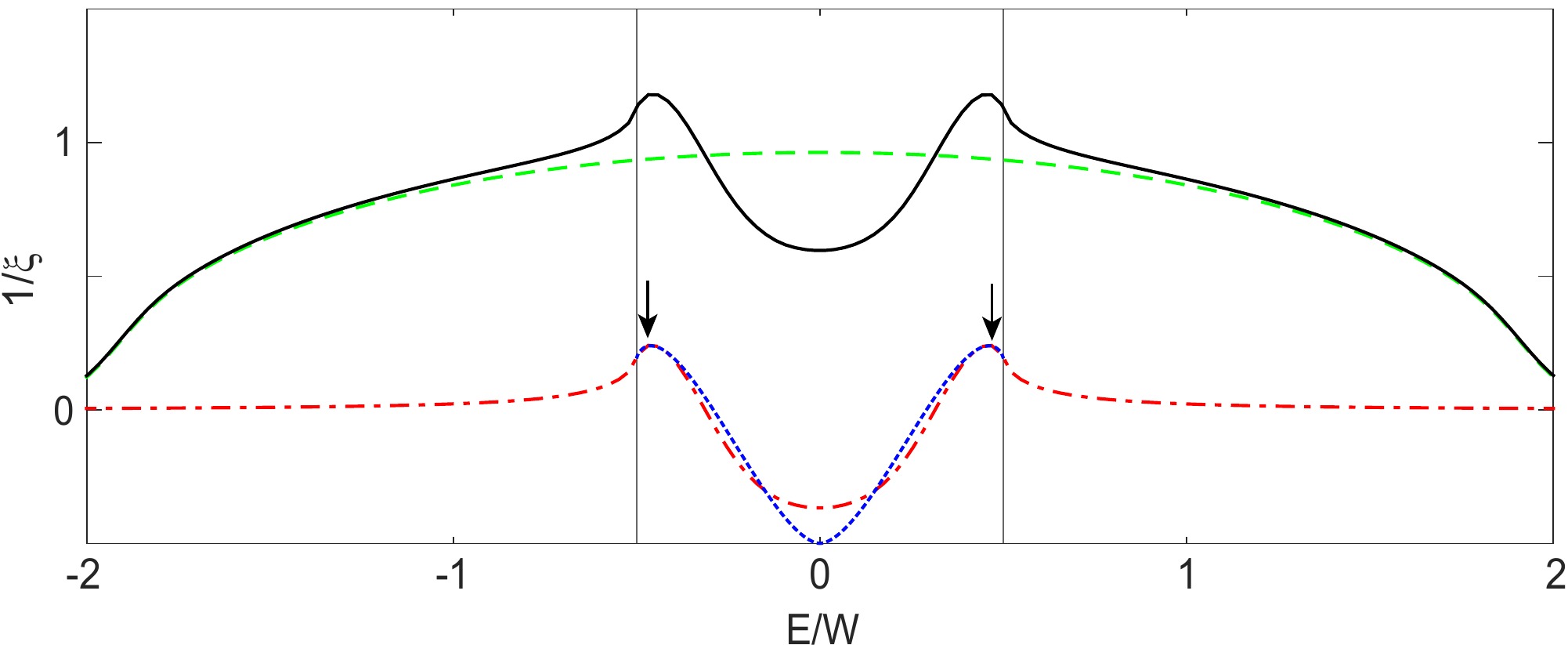}
\caption{ Inverse localization length of the Lieb lattice shown in Fig.~\ref{fig:bands}(b) with $W=0.5$ (thick solid line). The dash-dotted and dashed lines show the flat band and dispersive band term in Eq.~(\ref{eq:summation}), respectively. The dotted line shows the analytical approximation (\ref{eq:linearApprox}) in $|u|<1/2$ (delimited by the vertical lines), and the arrows mark its peak positions at $|u|=0.46$, which agree well with those of the numerical results.}\label{fig:lambda_decompose}
\end{figure}

This perturbation argument, however, cannot be applied to the localization length: even though $\rho_{DB}(z;W)$ is zero near the flat band energy for weak disorder, Eq.~(\ref{eq:dos}) tells us that $\xi_{DB}^{-1}(E;W)$ here is not necessarily small and hence cannot be neglected. As an example, we show $\xi^{-1}(E;W)$ and its two terms in Fig.~\ref{fig:lambda_decompose}. We find that $\xi_{DB}^{-1}(E;W)$ in fact has a larger magnitude than $\xi_{FB}^{-1}(u)$ in $|E|\lesssim W/2$, and the negligence of the former would lead to a negative and hence unphysical value of the localization length.
Nevertheless, we note that $\xi_{DB}^{-1}(E;W)$ is almost a constant for $|E|\lesssim W/2$ and weak disorder as Fig.~\ref{fig:lambda_decompose} shows. This property is due to the slowly varying kernel $\ln|E-z|$ in Eq.~(\ref{eq:dos}) and the vanishing of $\rho_{DB}(z;W)$ near the flat band energy for weak disorder (see Fig.~\ref{fig:dos}). Hence $\xi_{DB}^{-1}(E;W)$, which represents the coupling to the dispersive bands, does not play a role in the non-monotonic energy dependence of $\xi(E;W)$ near $E_{FB}$, and the anomalous minimum of $\xi(E;W)$ on each side of $E_{FB}$ is solely due to the flat band itself: as $\xi_{FB}^{-1}(E)$ in Fig.~\ref{fig:lambda_decompose} clearly shows, these minima correspond to the maxima of $\xi_{FB}^{-1}(E)$, and the latter can be captured by a crude (and linear) approximation for the DOS of the flat band, i.e., $\rho_{FB}(u) \approx 2-4|u|\,(|u|<1/2)$. Once substituted into Eq.~(\ref{eq:dos_lieb3}), it gives the following analytical expression
\begin{align}
\xi_{FB}^{-1}(u) \approx -\frac{1}{2} + |u|\ln\frac{\frac{1}{2}+|u|}{\frac{1}{2}-|u|}+2u^2\ln\frac{\frac{1}{4}-u^2}{u^2}\label{eq:linearApprox}
\end{align}
for $|u|<1/2$, which agrees reasonably well with its value obtained numerically (see Fig.~\ref{fig:lambda_decompose}), including its peak positions (and hence also the positions of the minima of the localization length) near $|u|=0.46$.

We further note that $\xi_{DB}^{-1}(E;W)$ is also insensitive to a weak disorder $W$ for a fixed $E$ (in addition to $E$ for a given $W$ mentioned above) near the flat band, again due to the slowly varying kernel $\ln|E-z|$ in Eq.~(\ref{eq:dos}). It smears out the change of $\rho_{DB}(E;W)$ when $W$ changes to $W'$, i.e.,
\begin{align}
\xi_{DB}^{-1}&(E;W) - \xi_{DB}^{-1}(E;W') \approx \int_{DB} \ln\left|\frac{E-z}{J}\right| \frac{dz}{\Gamma} \cdot \nonumber \\
&\cdot \int_{DB} [\rho_{DB}(z;W)-\rho_{DB}(z;W')]\,dz=0,
\end{align}
when the energy $E$ is far from the dispersive bands. The subscript ``$_{DB}$" of the integrations denotes integrating over the bandwidth $\Gamma$ of the dispersive bands, and the above difference vanishes due to the second integral and the same normalization of $\rho_{DB}(E;W),\rho_{DB}(E;W')$. The vanishing difference above explains why the different localization curves in Fig.~\ref{fig:loclen}(b) overlap in $|u|\lesssim 1/2$, which leads to a scaling behavior
\be
\xi(E;W) \approx \xi(u)\quad(|u|\lesssim 1/2)\label{eq:scaling}
\ee
for weak disorder. We note that the $W$-independence of the localization length right at $E=E_{FB}$ (and $u=0$) found previously \cite{Leykam,Flach,Baboux} is a special case of Eq.~(\ref{eq:scaling}).

\section{Flat band states at the localization minimum}

In the previous section we have established that the anomalous minimum of the localization length on each side of $E_{FB}$ is solely due to the DOS of the flat band and independent of the opposing band. In this section we discuss the spatial profile of the states at these localization minima, which we found to be the compact states that exist even in the absence of disorder \cite{Manninen2,Baboux}.

\begin{figure}[t]
\includegraphics[clip,width=\linewidth]{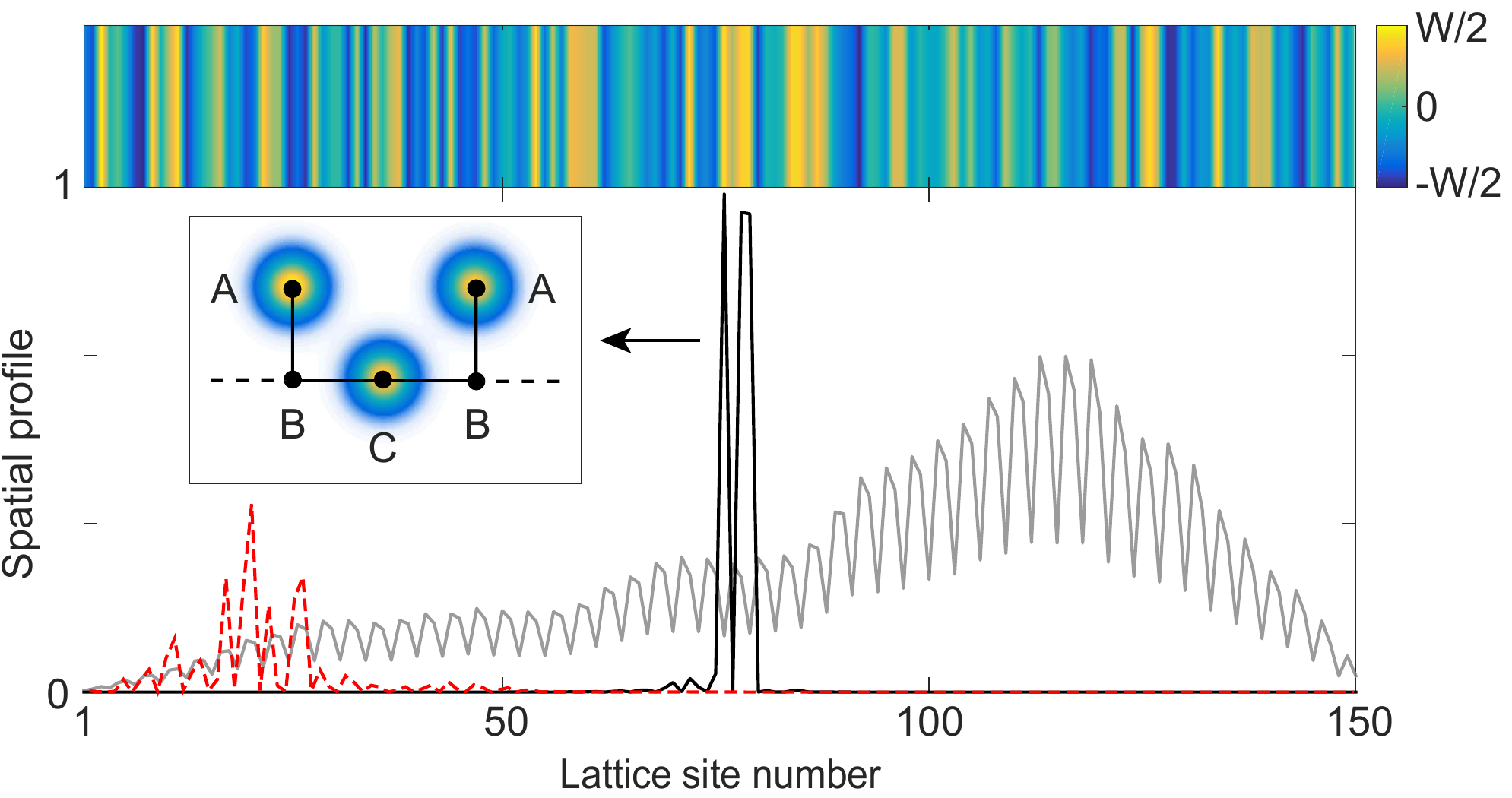}
\caption{ Top: Example of on-site disorder in a Lieb lattice with 150 sites (50 unit cells) and $W=0.5$. The value of the disorder on each lattice site is illustrated using the color scale on the right. Bottom: Spatial profiles of three representative states in this example, including a compact flat band state near $u=0.41$ (dark solid line), a more extended flat band state near $u=0.08$ (red dashed line), and a dispersive band state (grey solid line). Inset: Schematic of the compact state overlaid with the lattice structure.}\label{fig:wf}
\end{figure}

We have noted that $E=E_{FB}\pm W/2$ delimit the non-zero range of $\rho_{FB}(u)$ when discussing Fig.~\ref{fig:dos}. This observation is in fact quite counterintuitive but also instructive: the maximum energy shift of each lattice site is $W/2$, and in order to shift the energy of a state spanning several lattice sites by this amount, all these lattice sites then need to have the same disorder potential $W/2$. The probability of such a rare event is negligible for modes in the dispersive band, which have a long localization length (see the grey line in Fig.~\ref{fig:wf}). In the flat band however, there are compact states that span only two unit cells in the absence of disorder \cite{Manninen2,Baboux}, and they are ``dark" on the $B$ sites where their wave functions vanish due to the cancellation of hopping from neighboring $A$ and $C$ sites (see Fig.~\ref{fig:wf}; inset). Hence these compact states are insensitive to the disorder on the $B$ sites and only require three similar on-site disorder potentials on two consecutive $A$ sites and the $C$ site in between to have a nearly unaltered compact state. One example of a compact state at $u=0.41$ is shown in Fig.~\ref{fig:wf}, and we note that for a uniform disorder, the probability of having such compact states near $|u|=1/2$ is the same as that near $|u|=0$, which leads to a long tail of $\rho_{FB}(u)$ (up to $|u|=1/2$) when compared with that of $\rho_{DB}(E;W)$ inside the band gaps (see Fig.~\ref{fig:dos}).

These compact states are weakly coupled to the rest of the disordered lattice and lead to the minima of $\xi(u)$ near $|u|=1/2$: the flat band modes other than these compact states are more extended along the lattice, and they sample more on-site disorders which tend to cancel each other (see the flat band mode at $u=0.08$ shown by the dashed line in Fig.~\ref{fig:wf}). Therefore, the energies of these more extended flat band modes are closer to the undisturbed flat band energy, which leads to the increase of the localization length as $|u|$ approaches 0. In $|u|\geq1/2$, the flat band states do not exist and $\xi_{FB}^{-1}(u;W)$ approaches zero (see Fig.~\ref{fig:lambda_decompose}), and as a result, the localization length gradually increases to its value given by $\xi_{DB}(E;W)$ due to the dispersive bands.

\begin{figure}[t]
\includegraphics[clip,width=\linewidth]{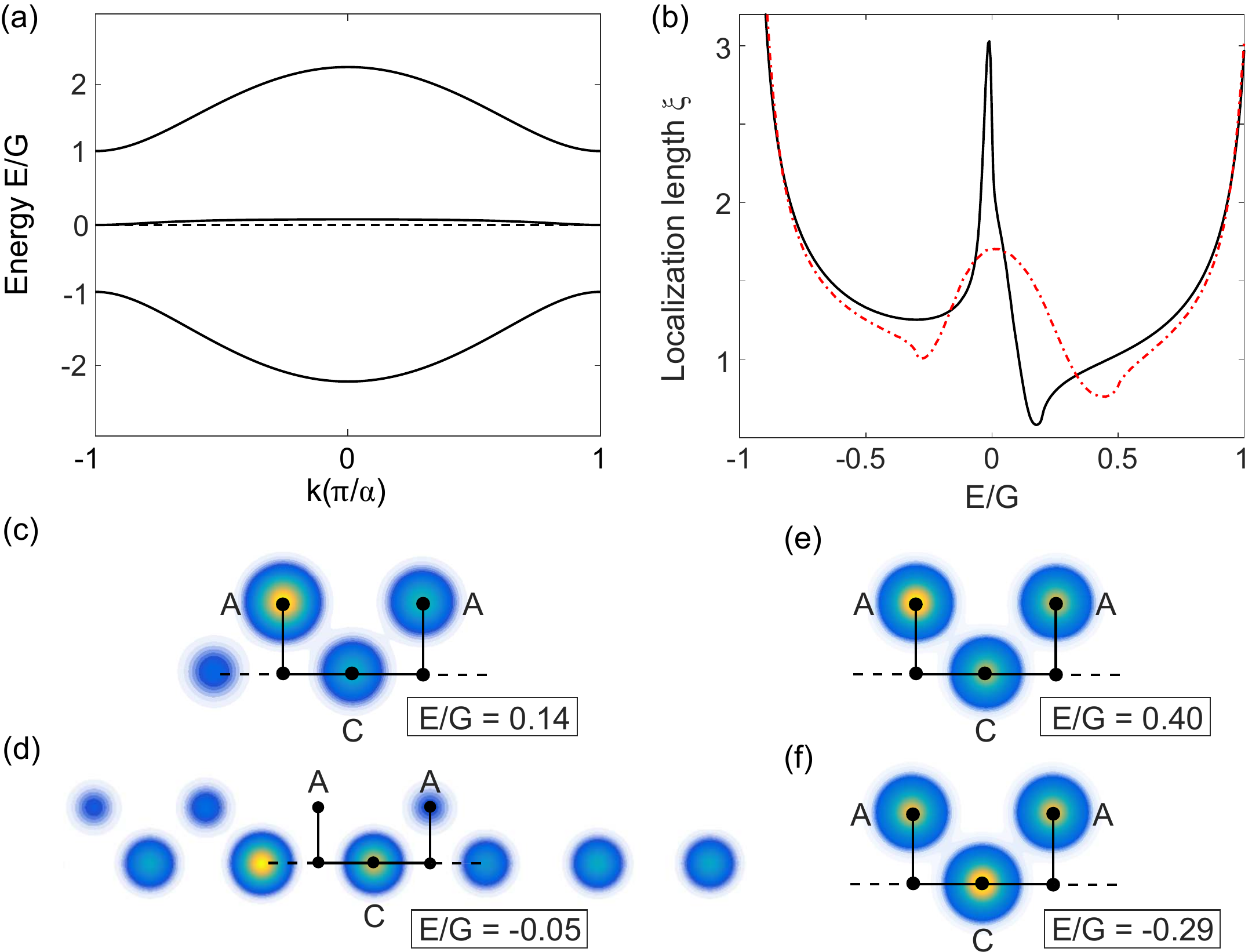}
\caption{ Band structure (a) and localization length (b) of a 1D Lieb lattice with a bent band in the middle. In (a) all the parameters are the same as in Fig.~\ref{fig:bands}(b) except for $E_A=0.1$. The center band has $\max(E)=0.08$ and $\min(E)=0$, and the dashed line shows the flat band when $E_A=0$. Its localization length in (b) is calculated with $W=0.2$ (solid line) and 0.8 (dash-dotted line). (c) and (d) Typical modes on the high and low energy sides of the bent band with $W=0.2$, respectively. (e) and (f) Same as (c) and (d) but with $W=0.8$.}\label{fig:bend}
\end{figure}

To further emphasize the role of the compact states in forming the anomalous minima of the localization length near $E_{FB}$, we note that these states persist even when the flat band is slightly bent, for example, by introducing a small detuning $\varepsilon>0$ to the $A$ sites [see Fig.~\ref{fig:bend}(a)]. Note however, they now only appear on the high energy side of the bent band when $W/2\lesssim \varepsilon$ [see Figs.~\ref{fig:bend}(c) and (d)], and hence the anomalous minimum of the localization length should only appear on the high energy side also, as we verify in Fig.~\ref{fig:bend}(b); the minimum of the localization length on the low energy side of the bent band, in contrast, follows the conventional scenario due to the lack of the compact states, and the curvature of the localization length does not change its sign inside this band gap. As the disorder strength increases, i.e., when $W/2$ becomes large when compared with $\varepsilon$, the compact states appear on both sides of the bent band [see Figs.~\ref{fig:bend}(e) and (f)], and the anomalous minimum of $\xi(E;W)$ now reappears on the low energy side of the bent band [see Fig.~\ref{fig:bend}(b)].

\section{Conclusion and Discussion}

In summary, we have shown that anomalous localization minima near an isolated flat band is a generic feature caused by its sharp DOS. In the previous instance of these minima found in a photonic crystal waveguide, they exist for evanescent waves in a defect-free system, where the flat band is due to a Van Hove saddle point singularity. Here we have shown that they exist in the presence of disorder in a Lieb lattice, where the flat band is the result of canceled quantum tunneling from $A$, $C$ sites to the neighboring $B$ site.

As a result of this generic property, when a more exotic band gap is formed by two flat bands, we find two such localization minima inside this band gap, which highlights their anomaly in view of the conventional midgap minimum of the localization length exemplified in Fig.~\ref{fig:loclen}(a). These anomalous minima are direct manifestations of the compact states in the flat band, which persist even when the flat band is slightly bent. In the examples we have shown the on-site energy is subjected to a uniform disorder, and we have checked that these features hold for other types of disorder as well, such as a Gaussian disorder (see Appendix C).

As mentioned previously, the tight-binding model given by Eqs. (\ref{eq:A0})--(\ref{eq:C0}) can be applied to different types of particles as long as the noninteracting picture applies. Therefore, although the variation of the localization length near $E_{FB}$ is on the order of the lattice constant and hence difficult to observe for electrons, it can be probed in synthesized systems as numerically shown in Refs.~\citenum{PhC1} and \citenum{PhC2}, using a quasi-1D dielectric photonic crystal waveguide. The candidates to observe these anomalous localization minima on a Lieb lattice include cold atoms in optical lattices and exciton-polariton condensates in micropillar cavities \cite{Baboux,Jacqmin}, as long as the interaction between the underlying particles is kept small. These systems have a lattice constant on the order of microns \cite{Baboux,Jacqmin} or tens of microns \cite{Vicencio,Mukherjee,Silva}, and hence they are ideal for verifying the anomalous minima of the localization length (see the experimental technique used in Ref.~\citenum{Baboux}, for example).

We have also shown that the localization length displays a scaling behavior close to $E_{FB}$, which holds even when the flat band is shifted toward one dispersive band (see Appendix \ref{sec:compare}). Using the top right element of the Green's matrix, we have derived an analytical relation between the DOS and the localization length for the 1D Lieb lattice. It takes the form of a summation for the inverse localization length, with one flat band term and one dispersive band term, and we have used it to explain the localization properties mentioned above.

Finally, we note that while localization in 1D disordered systems is qualitatively different from 2D and three-dimensional systems in general \cite{Lee}, the anomalous minima of the localization length near an isolated flat band are tied to the compact states which exist in higher dimensions as well \cite{Manninen2,Flach}.  Hence it is likely that these anomalous minima may exist beyond 1D, even though they have not been found in 2D systems with a flat band touching a dispersive band \cite{Chalker}.

\appendix
\section{Transfer matrix calculation}
\label{sec:Tmatrix}
The transfer matrix method is a convenient way to calculate the localization length in 1D \cite{Kramer}. It relies on the following relationship between the values of the wave function on three consecutive lattice sites of the same type:
\be
X_{j+1} = d_j X_{j} + e_{j-1} X_{j-1}.\label{eq:itr1}
\ee
The transfer matrix $\bm{\tau}_j$ for neighboring unit cells is then defined by
\be
\begin{pmatrix}
X_{j+1}\\Xj
\end{pmatrix}
=
\begin{pmatrix}
d_j & e_{j-1}\\
1 & 0
\end{pmatrix}
\begin{pmatrix}
Xj \\ X_{j-1}
\end{pmatrix}
\equiv
\bm{\tau}_j
\begin{pmatrix}
Xj \\ X_{j-1}
\end{pmatrix},
\ee
and the total transfer matrix is given by
\be
\bm{T} =  \prod^{N-1}_{j=2} \bm{\tau}_j,\label{eq:Tmatrix}
\ee
satisfying
\be
\begin{pmatrix}
X_N\\X_{N-1}
\end{pmatrix}
=
\bm{T}
\begin{pmatrix}
X_2 \\ X_1
\end{pmatrix}.
\ee
The exponent of its largest eigenvalue $\lambda_\text{max}$, after ensemble average (denoted by $\langle\cdot\rangle$) and divided by $N$,
gives the inverse localization length $\xi^{-1}$ in the asymptotic limit $N\rightarrow\infty$:
\be
\xi^{-1} = \lim_{N\rightarrow\infty}\frac{1}{N}\langle\ln(\lambda_\text{max})\rangle. \label{eq:loc}
\ee

\begin{figure}[h]
\centering
\includegraphics[clip,width=0.6\linewidth]{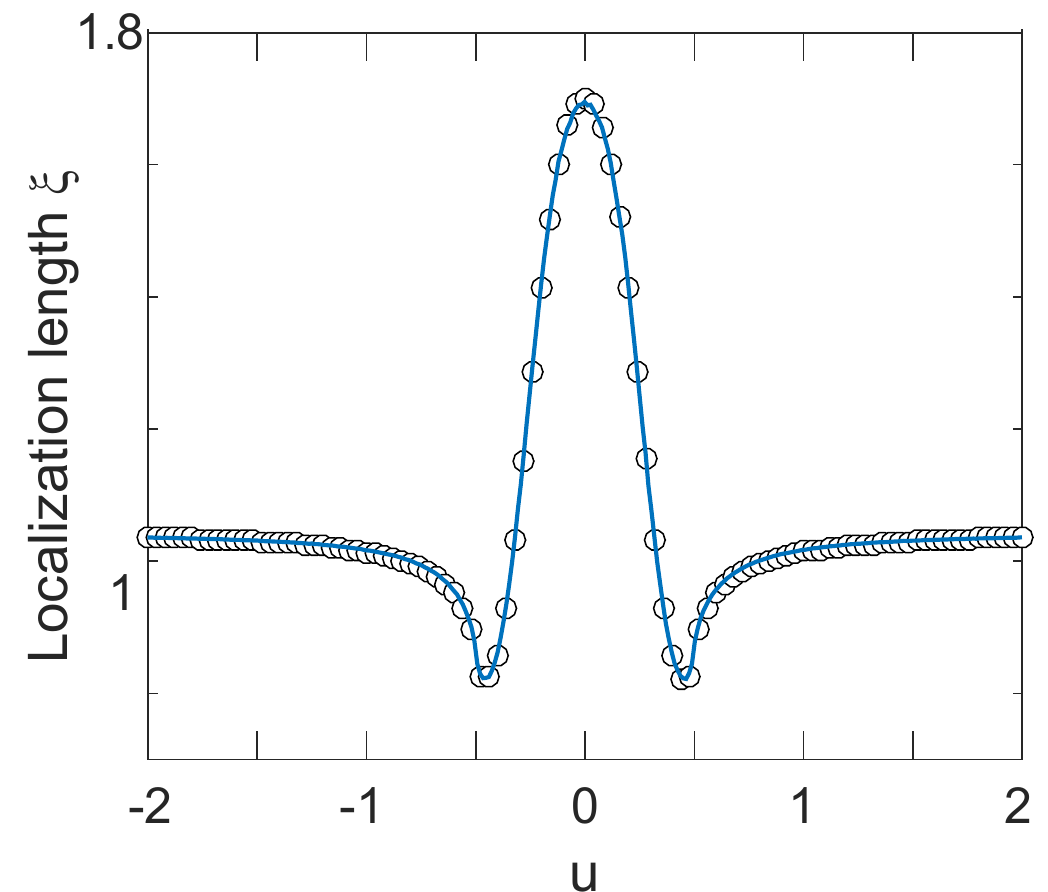}
\caption{ Localization length in a 1D Lieb lattice at $W=0.05$. The solid line and open circles are calculated using the recursive relation (\ref{eq:itr2}) and (\ref{eq:itr4}), respectively. The other parameters are the same as in Fig.~\ref{fig:bands}(b) in the main text.}\label{fig:b_vs_c}
\end{figure}

The recursive relation (\ref{eq:itr1}) can be easily found for $A$, $B$, or $C$ sites of the 1D Lieb lattice shown in Fig.~\ref{fig:bands}(b) of the main text. Take $B$ sites for example, by expressing $A_j$, $C_j$ in terms of $B_j$ and $B_{j+1}$ and plugging them into Eq.~(\ref{eq:B0}) in the main text, we find
\begin{align}
&\hspace{6mm} f_j B_j =   \chi_j  B_{j+1} + \chi_{j-1} B_{j-1},\label{eq:itr2}\\
f_j &= \Delta_{Bj} - \frac{G^2}{\Delta_{Aj}}  - \chi_j  - \chi_{j-1},\quad \chi_j = \frac{J^2}{\Delta_{Cj}},
\end{align}
or equivalently $d_j=f_j/\chi_j, e_{j-1}=-\chi_{j-1}/\chi_j$ in the form of Eq.~(\ref{eq:itr1}). This is the option we have chosen in the main text.

Similarly, we find the following recursive relation for $C$ sites
\begin{align}
&\hspace{6mm}f_j C_j = \chi_{j+1} C_{j+1} + \chi_j C_{j-1},\label{eq:itr4}\\
f_j &= \Delta_{Cj} - \chi_j - \chi_{j+1},\quad
\chi_j = \frac{J^2}{\Delta_{Bj}-\frac{G^2}{\Delta_{Aj}}},
\end{align}
which gives the same localization length as the first choice above (see Fig.~\ref{fig:b_vs_c}).

\section{Compare localization length obtained using different methods}
\label{sec:compare}

\begin{figure}[h]
\includegraphics[clip,width=\linewidth]{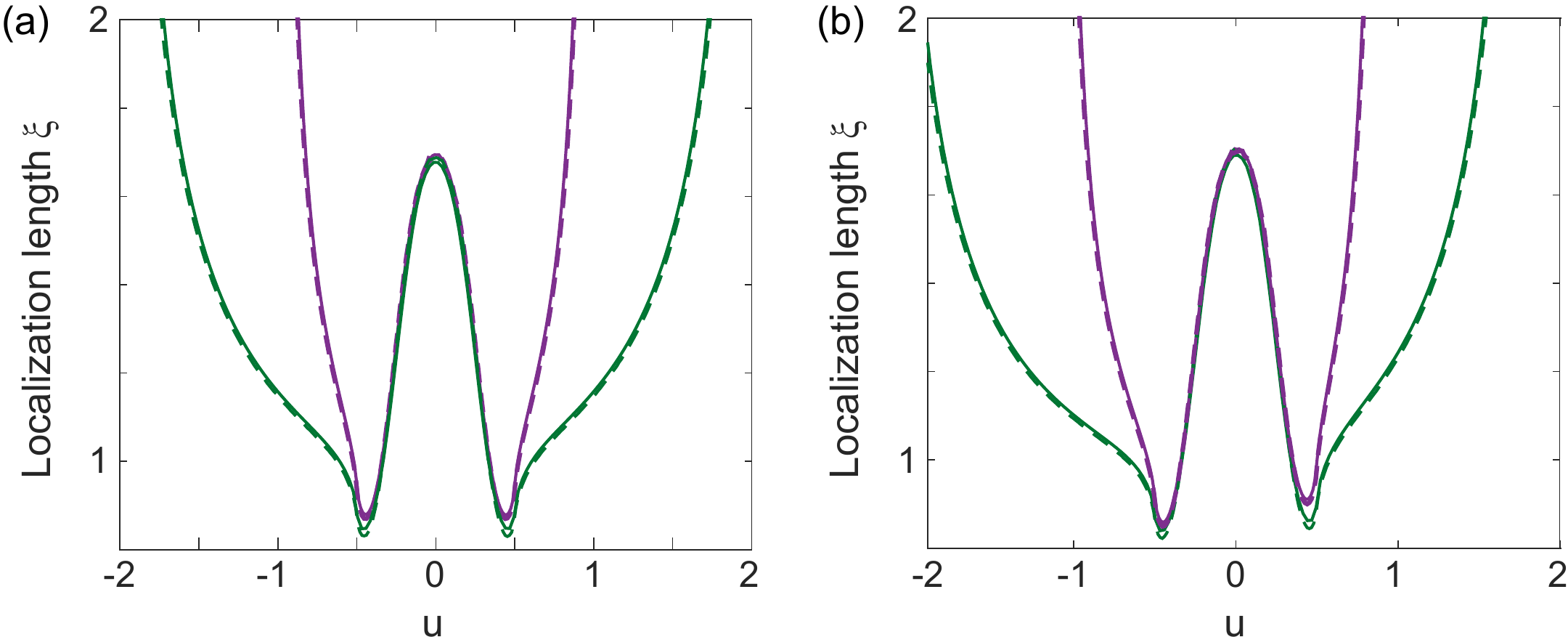}
\caption{ (a) Localization length of the Lieb lattice shown in Fig.~\ref{fig:loclen}(b) of the main text (solid lines), calculated with $W=0.5$ (bottom curve) and 1 (top curve) using the transfer matrix method. The dashed lines show the results using Eq.~(\ref{eq:summation}) in the main text and numerically calculated DOS (see Fig.~\ref{fig:dos} for example). (b) Same as (a) but with the flat band shifted to $E_{FB}=0.1$.
}\label{fig:loc_lieb_dos}
\end{figure}

The results obtained from the transfer matrix and the analytical expression (\ref{eq:summation}) in the main text agree well, which we illustrate in Fig.~\ref{fig:loc_lieb_dos}. In Fig.~\ref{fig:loc_lieb_dos}(a) the flat band sits at $E_{FB}=0$ as in Fig.~\ref{fig:bands}(a) of the main text, and in Fig.~\ref{fig:loc_lieb_dos}(b) we shift the flat band to $E_{FB}=0.1\equiv\varepsilon$ without changing the dispersive bands. The latter is achieved by choosing $E_A=E_C=\varepsilon$, $E_B=-\varepsilon$ and $G\rightarrow \sqrt{G^2-\varepsilon^2}$. We note that the scaling behavior given by Eq.~(\ref{eq:scaling}) in the main text holds near $E_{FB}$ even though $E_{FB}$ is shifted. This result is consistent with the proof of Eq.~(\ref{eq:scaling}) we have given, and all it requires is that the dispersive band contribution $\xi_{DB}^{-1}(E;W)$ is insensitive to $W$, which is satisfied when the shifted $E_{FB}$ is still far from the dispersive band edges with weak disorder. We do note that $\xi(E;W)$ now shows a slight asymmetry about $E=E_{FB}\neq0$ ($u=0$), which comes from the contribution of the two dispersive bands, which are no longer symmetric with respect to the shifted flat band.

\section{Localization length with a Gaussian disorder}
\label{sec:Gaussian}

\begin{figure}[b]
\includegraphics[clip,width=\linewidth]{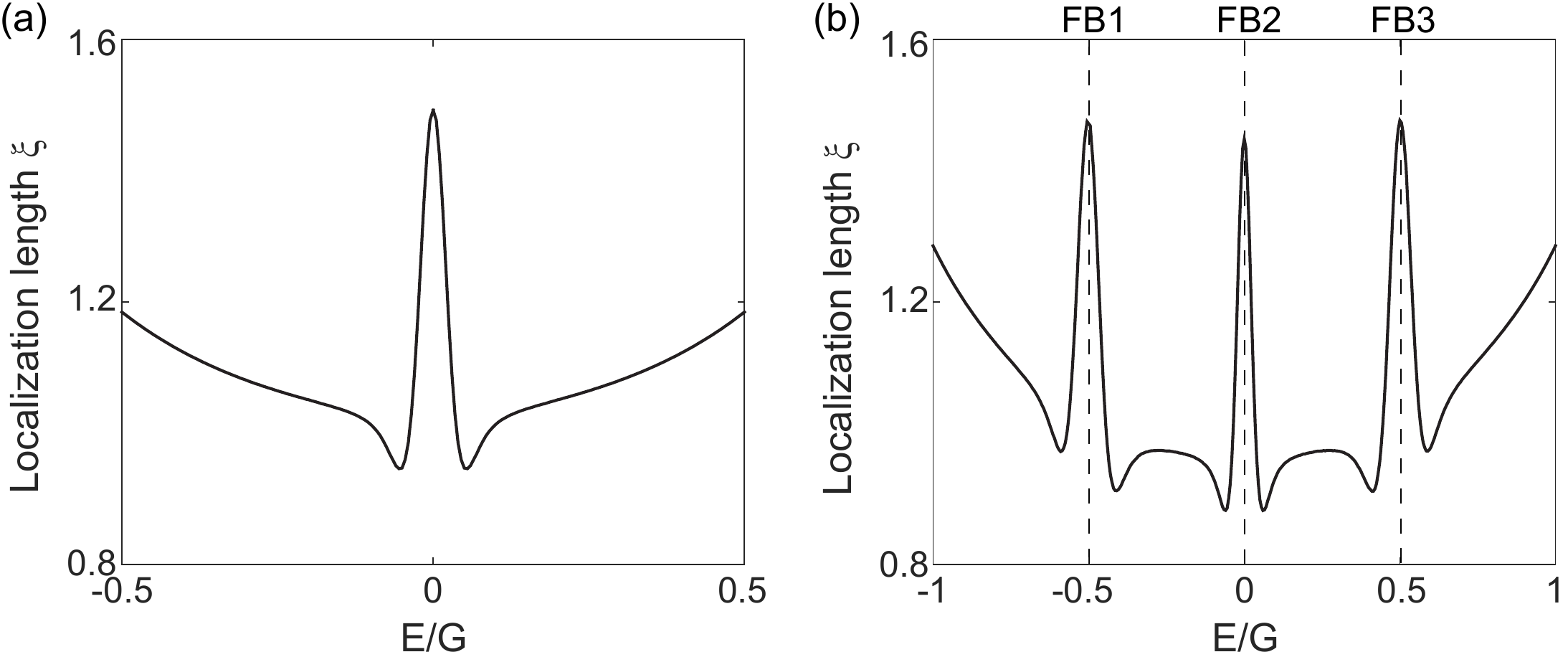}
\caption{Anomalous minima of the localization length with a Gaussian disorder, for (a) the Lieb lattice in Fig.~\ref{fig:bands}(b) and (b) the double-sided Lieb lattice in Fig.~\ref{fig:2minima}. We express the variance of the on-site disorder in the same form as that for a uniform disorder, i.e., $\langle E^2_{X_j}\rangle=W^2/12$, where $W=0.1$ in (a) and 1/6 in (b).}\label{fig:Gaussian}
\end{figure}

In the conclusion section of the main text, we mentioned that the anomalous minima of the localization length near an isolated flat band occur for other types of disorder as well. Here we show two examples using a Gaussian disorder, one for the Lieb lattice in Fig.~\ref{fig:bands}(b) and one for the double-sided Lieb lattice in Fig.~\ref{fig:2minima}. The results are shown in Fig.~\ref{fig:Gaussian}, where two and six such minima are clearly seen in these two cases, respectively.

\noindent \textbf{Acknowledgements.} The author thank Hakan T\"ureci, Vadim Oganesyan, Matteo Biondi, Sebastian Schmidt and Bo Zhen for stimulating discussions. This project is supported by NSF under Grant No. DMR-1506987.


\end{document}